\documentclass[aps, amsmath,12pt]{revtex4}

\usepackage{dsfont}
\usepackage{natbib}
\usepackage{amssymb}
\usepackage{graphicx}
\usepackage{amsmath}
\usepackage{url}
\usepackage{epstopdf}
\usepackage{color}
\usepackage{feynmp}
\usepackage{grffile}
\usepackage{bbold}
\usepackage{soul}

\def\be {\begin{equation}}
\def\ee  {\end{equation}}
\def\bea {\begin{eqnarray}}
\def\eea {\end{eqnarray}}

\begin{document}

\title{ADM Energy and Infra-gravitons}

\author{A. P. Balachandran}
\email{balachandran38@gmail.com}
\affiliation{ Physics Department, Syracuse University, Syracuse, New York 13244-1130, U.S.A.} 
\affiliation{Chennai Mathematical Institute, Kelambakkam 603103, India}

\author{Babar Qureshi}
\email{ babar.qureshi@lums.edu.pk}
\affiliation{Department of Physics, Lahore University of Management Sciences, Lahore, Pakistan} 

\date{19/12/2015}

\begin{abstract}

In QED and QCD~\cite{Balachandran:2015dva, Balachandran:2014hra}, infrared photons and gluons alter the Hilbert space of the theory so that the in and out states live in a Hilbert space carrying a representation of basic commutation relations which is non-equivalent to the standard Fock representation, leading to spontaneous breakdown of Lorentz symmetry and physical effects. These results are generalized here to quantum gravity regarded as an $SL(2,\mathcal{C})$ gauge theory and the shift of the ADM energy density and mass from infrared spin connection (the spin cloud) is discussed.

\end{abstract}

% insert suggested PACS numbers in braces on next line
% insert suggested keywords - APS authors don't need to do this
%\keywords{}

\maketitle

 \section{Introduction}
In QED, infrared photons dress the charged particles such as the electrons and alter its interacting in and out states. The particle thus gets a form factor which makes its mass and spin shells fuzzy. The resulting Hilbert space has a unitarily non-equivalent representation of the basic field commutation relations compared to the standard Fock representation: a physical result of which is that Lorentz group can not be implemented unitarily on the in and out states.

These results are due to Roepstorff\cite{Roep}, Buchholz et al.\cite{Buchhold} and  especially Fr{\"o}hlich et al.~\cite{Frohlich}. A canonical fixed time approach to these results was developed in ~\cite{Balachandran:2014hra} where their experimental implications were also examined. They were adapted in ~\cite{Balachandran:2015dva} to QCD and non-abelian gauge theories with similar conclusions. Color symmetry was also shown to be spontaneously broken just like Lorentz invariance.

These results are here generalized to quantum gravity formulated as an $SL(2,\mathcal{C})$ gauge theory of spin connection. Infra-gravitons ( or rather infra-$SL(2,\mathcal C)$ gluons or the `spin cloud') then shift the ADM energy density which is estimated. 

For previous treatments of gravitational infra-red effects see \cite{Alvegard:1978dq,Ware:2013zja} and references therein. 

\section{The ADM Mass} The ADM Poincar\'e group is defined on asymptotically flat spacetimes which are the spacetimes considered in this paper. The ADM energy is then written as a surface integral at spatial infinity. If $g$ is the spacetime metric and $\eta$ the Minkowski metric, then on spatial slice $\mathbb{ M}^3$,
\begin{eqnarray}
&&g_{\mu\nu}=\eta_{\mu\nu}+h_{\mu\nu},\label{metricasymp}\\
&&h_{\mu\nu}=\mathcal{O}(\frac{1}{r})\ \ \ \ \textrm{as}\ r\longrightarrow\infty,\nonumber
\end{eqnarray}
where $r$ is a radial coordinate defined in the asymptotic region ( complement of a ball $\mathbb B^3$ in $\mathbb M^3$). The ADM energy $E_{ADM}$ is then~\cite{Witten:1981mf}
\begin{equation}
E_{ADM}=\frac{1}{16\pi G}\lim_{r\longrightarrow \infty} \int_{S^2_r} d\Omega\, r^2\,\hat x^j\,(\partial_k g_{jk}-\partial_j g_{kk}).\label{EADM}
\end{equation}

Witten has written this expression using Dirac spinors and the self-dual spin connection $A$: 
\begin{equation}
E_W = E_{ADM}=\frac{i}{2\pi G}\lim_{r\longrightarrow \infty} \int_{S^2_\infty} \bar{\psi} \gamma_5\gamma\wedge A\psi,
\label{EADM2}
\end{equation}
\begin{equation}
\psi=\textrm{constant on } S^2_\infty,\ \ \psi^\dagger\psi=1.
\end{equation}

\section{ On Self-Dual Gravity} In view of (\ref{EADM2}), we use the approach of self-dual gravity~\cite{lqg} where the basic configuration space variables are the spatial parts of the spin connection $A_i^{\alpha\beta}$ which are ``self-dual'':
$$A_i^{\alpha\beta}=\frac{i}{2}\epsilon^{\alpha\beta}_{\phantom{\alpha}\phantom{\beta}\gamma\delta}\,A_i^{\gamma\delta}.$$
Hence $A_i^{jk}$ and $A_i^{0l}$ ($j,k,l\in [1,2,3])$) are related:
$$A_i^{0\rho}=\frac{i}{2}\epsilon_i^{\rho mn}A_{i,mn},\ \ \rho=1,2,3.$$
It follows also that $A_i$ is not hermitian. The non-hermitian part of $A_i$ is given by the extrinsic curvature. The connection entering the Witten equation ~\cite{Witten:1981mf} $i\gamma^lD_l\psi=0$ is this connection. It is not the connection compatible with the spatial metric {\it q}.

In the self-dual approach, the conjugate field $E_\sigma^l$ to $$A_i^{0,\rho}\colon=A_i^\rho,$$
\begin{equation}
[A_i^{\rho}(x)\, ,\, E_\sigma^j(y)]=i\delta^\rho_\sigma\delta^j_i\delta^3(x,y)\label{ccr}
\end{equation}
on spatial slice is given by the frame field $e_i$:
$$E^i_\sigma=\frac{1}{2}\epsilon_{\sigma\rho\nu}\epsilon^{ijk}e^\rho_j e^\nu_k.$$
Here $e_j^{\rho}$ are frames and the gauge choice $e_i^0=0$ is being made. Also, all the indices in this equation run from 1 to 3. 

Since
$$E^i_\sigma(r)e^\sigma_l(r)=|e|\delta^i_l ,$$
where $|M|=\textrm{det}\ M,$
we get $$|E|=|e|^2$$
 and 
\begin{equation}
[A_i^\alpha(x),e_j^\beta(y)]=\frac{i}{|e|}[\frac{1}{2}e_i^\alpha(x)e_j^\beta(x)-e^\beta_i(x)e^\alpha_j(x)]\delta^3(x,y)\label{Aecomm}
\end{equation}
on the spatial slice.

This expression asymptotically simplifies on using
{\begin{equation}
e_i^\alpha\sim\delta_i^\alpha+\mathcal{O}(1/r)\label{frameasymp}
\end{equation}
to the following expression
\begin{equation} \label{Aecomm}
[A_i^\alpha(x),e_j^\beta(y)]=i[\frac{1}{2}\delta_i^\alpha\delta_j^\beta-\delta^\beta_i\delta^\alpha_j]\delta^3(x,y)+\mathcal{O}(\frac{1}{r}).
\end{equation}

{\it Remarks:}

The paper uses the strong resemblance of self-dual gravity and Yang-Mills theory. This resemblance has been noted before in the literature. A particular point worthy of emphasis and not discussed sufficiently in literature is the fact that there are global charges associated with Gauss law in both theories.

Thus in Yang-Mills theories with no matter sources, the Gauss law is $D^iE_i\approx 0$ where $E_i,A_j$ are conjugate variables as in eq. (\ref{ccr}). Since $E_i$ are distributions in quantum theory, we should smear it with Lie algebra valued test functions $\Lambda =\Lambda ^{\alpha}\lambda_\alpha$ ($\{\lambda_\alpha\}=$ a basis of Lie algebra), where $\Lambda^\alpha$ are of compact support, and following Dirac, write 
\begin{equation}
\mathcal G(\Lambda)|\cdot\rangle:=\int (D^i\Lambda^\alpha)E_{i,\alpha}|\cdot\rangle=0.
\end{equation}
Compactness of support of $\Lambda^{\alpha} $ lets us partially integrate and formally recover the standard form of Gauss law
\begin{equation}
\int\Lambda^\alpha (D^iE_i)_\alpha |\cdot\rangle=0.
\end{equation}
If now we relax the support condition on $\Lambda^\alpha$ and allow test functions $\chi^\alpha$ going to constants at infinity, we get
$$
Q(\chi)=\int (D^i\chi_i)_\alpha(E_i)^\alpha, \ \ \ \ \chi=\chi^\alpha\lambda_\alpha
$$ which however need not vanish on states.

All observables commute with $\mathcal G(\Lambda)$ and hence also with $Q(\chi)$.

Modulo the constraints, $Q(\chi)$ obey the Lie algebra of the gauge group,
\begin{equation}
[Q(\chi_1),Q(\chi_2)]=Q([\chi_1,\chi_2])
\end{equation}
and are the global charges. 

All these points are discussed in \cite{Balachandran:2014voa} and references therein.

In self-dual gravity, since the Gauss law constraint holds, classically, all the above considerations hold true. The global surviving group is $SL (2, \mathbb C)$. The meaning of this group for quantum gravity is not understood.

\section{ ADM Energy: The Calculation}  Our focus is on the asymptotic region of quantum gravity where the relations (\ref{metricasymp},\ref{frameasymp}) hold.

In QED and QCD, the asymptotic radiation field dresses the particle Fock state of momentum {\it P} to give the dressed in state. We can calculate it by adding the interaction
$$e\int\,d^3x\, A_\mu j^\mu(x)$$
in QED, where $j^\mu$ is the particle current \cite{Balachandran:2015dva}:
\begin{equation}
j^\mu(x)=\int\, d\tau\delta^4(x-z(\tau))\frac{dz^\mu(\tau)}{d\tau}.
\end{equation}

As we are considering infrared radiation, for a particle of mass $M$, we can set 
\begin{equation}
\frac{dz^\mu(\tau)}{d\tau}=\frac{P^\mu}{M},\ z(\tau)=\tau\frac{P}{M},\ \tau=t
\end{equation}
in which case the state is 
\begin{equation}
|P,\cdot\rangle_{\textrm{in}}=V(j)|P\rangle|0\rangle_\gamma
\end{equation}
where $|P\rangle$ and $|0\rangle_\gamma$ are the charged particle  and photon vacuum states respectively and
\begin{equation}
V(j)=T \exp\left\{-i\int _{\infty<t\leq 0}\,d^4x A_\mu(x)j^\mu(x)\right\}
\end{equation}
with the $t$ integral limited as shown. For details see \cite{Balachandran:2015dva}.

In QCD, $J^\mu$ is the current of a Wong particle \cite{Balachandran:2015dva}. It has ``internal'' degrees of freedom $$\hat{\lambda}=(\hat\lambda_1(\tau), \hat\lambda_2(\tau),\cdots,\hat\lambda_8(\tau))$$ which represent the Lie algebra of the gauged group $G$ with structure constants $c_{\alpha\beta}^\gamma$ on the particle states:
\begin{equation}
[\hat\lambda_\alpha,\hat\lambda_\beta]=ic_{\alpha\beta}^{\phantom{\alpha}\phantom{\beta}\gamma}\,\hat\lambda_\gamma.
\end{equation}
As shown in \cite{Balachandran:2015dva} , we can choose a gauge where $J^\mu$ has components only in the direction of an element $K$ in the Cartan subalgebra of $\underline{G}$:
\begin{equation}
J^\mu(x)=\hat K\, \int\, d\tau\delta^4(x-z(\tau))\frac{dz^\mu(\tau)}{d\tau}.
\end{equation}
The interaction now becomes
$$g\hat K\,\int\, d^3x[\mathrm{Tr} KA_\mu(x)]j^\mu(x),$$ where $g$ is the non-abelian coupling constant.

Choosing a particle state with eigenvalue $\lambda_j$ for $\hat K$, the in state is then seen to be 
\begin{eqnarray}
&&|P,\lambda_j,\cdot\rangle_{\textrm{in}}=V(J)|P,\lambda_j,\cdot\rangle\,|\textrm{Gluon vacuum}\rangle\\
&&V(J)=T\exp\left\{ -ig\,\lambda_j\int_{\infty<t\leq 0}\,d^4x\mathrm{Tr} (KA_\mu(x))j^\mu(x)\right\}. \label{instate}
\end{eqnarray}

In this gauge, there is matter source for $A_\mu$ only in the direction $K$. We can calculate how the particle state gets dressed by infra-gluons.

We now adapt the previous considerations to the quantum gravity case. We use it only in the asymptotically flat region where spacetime is Minkowskian. Also, the connection $A$ couples to the Dirac operator of matter in the standard way: it is $-i\gamma^{\mu}D_{\mu}$ where $D_{\mu}$ is the covariant derivative for connection $A$. In the asymptotic region, this is the Minkowskian Dirac operator for $SL(2,\mathcal{C})$ connection.

However, there is a problem as $SL(2, \mathcal C)$ is not compact. For this reason, any basis of the Cartan subalgebra contains a generator which generates the non-compact $\mathbb{R}^1$,  besides a generator which can be of the compact $U(1)$. The non-compact $\mathbb{R}^1$ leads to reality and unitarity problems as one can infer from the literature on Loop Quantum Gravity \cite{lqg}. We avoid these problems by restricting to radiation associated with the compact $U(1)$. Hence, for us, $\hat K $ is the generator of a compact $SU(2)$ rotation in $SL(2,\mathcal C)$. In other words, we assume that the internal particle states are induced from the representations of the $U(1)$ subgroup with generator $\hat K$. The eigenvalue of $\hat K$ in this representation is $\lambda_j$ assumed non-zero :
$$\lambda_j\in\{\pm1/2,\pm3/2,\cdots\}.$$
From $E_{ADM}$, we see that the shifted energy and its density can be obtained if we know
\begin{eqnarray}
&&a)\ \ V(J)^\dagger A_i V(J) \label{VAV}\\ 
&&b)\ \ V(J)^\dagger e_i V(J) \label{VeV}
\end{eqnarray}
in the asymptotic region.

{\it Remarks:}

The non-compact nature of $SL(2, \mathbb C)$ may in future developments be properly treated to formulate a consistent way to quantize gravity. Such a development is not likely to invalidate our results, but only modify them by additional contributions.

Ware et. al. \cite{Ware:2013zja} have also considered similar dressing by coherent  infra-gravitons. They consider perturbative linearized gravity action while our approach is fully non-perturbative.

\subsection{ Calculation of (\ref{VAV})}

We will see that this is the significant term.

We can find the commutator of $A_i$ with $A_j$ for different spacetime points using the leading term in the gradient expansion. That is appropriate for our extreme low energy problem.

The leading term of interest in the gradient expansion in the action is
\begin{eqnarray}
&&S_{\mathrm{eff}}=-\frac{\Lambda}{4}\int\,d^4x\,F^{K\,\mu\nu}(x)F^K_{\mu\nu}(x)\\
&&F_{\mu\nu}^K=\partial_\mu A^K_\nu-\partial_\nu A^K_\mu
\end{eqnarray}
where $A^K_\mu={\textrm tr} K A_\mu$ is the component of $A_\mu$ in the compact direction $K$ (see (\ref{instate})) and $\Lambda$ is an unknown constant. In the Lorentz gauge, it gives the field equation
\begin{equation}
\partial^\mu\partial_\mu A^K_\nu=0.
\end{equation}

The constant $\Lambda$ and the action $S_{\mathrm{eff}}$ must be emergent from quantum gravity. The one-loop diagram 
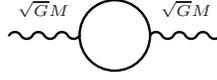
\begin{center}
\begin{fmffile}{sunset2}
\begin{fmfgraph*}(80,40)
\fmfleft{i}
\fmfright{o}
\fmf{wiggly}{i,v1}
\fmf{wiggly}{v2,o}
\fmf{plain,left,tension=0.5}{v1,v2,v1}
\fmfv{l={\tiny$\sqrt{G}M$},l.a=120,l.d=.1w}{v1}
\fmfv{l={\tiny$\sqrt{G}M$},l.a=60,l.d=.1w}{v2}
\end{fmfgraph*}
\end{fmffile}
\end{center}
from a Dirac field of mass $M$ induces this term, with $\sqrt{G}M$ replacing the electric charge $e$ in QED (\cite{Nair} and Sec. 7.5 of \cite{Peskin}). Note that we are interested only in the infrared limit of this diagram. We thus estimate that 
\begin{equation}
\Lambda=\xi(\sqrt G M)^2
\end{equation}
where we can only say that $\xi\neq 0$.

With $A'^K=\sqrt{\xi G} MA ^K$,
\begin{equation}
S_{\mathrm{eff}}=-\frac{1}{4}\int d^4x[\partial_\mu {A'^K}_\nu(x)-\partial_\nu {A'^K}_\mu(x)]^2.
\end{equation}

In the Lorentz gauge, ${A'^K}$ has the commutator

\begin{eqnarray}
&&[A'^K_\mu(x),A'^K_\nu(y)]=\eta_{\mu\nu}\, D(x-y),\\
&&D(x-y)=\int \frac{d^3p}{(2\pi)^3}\frac{1}{2p_0}[e^{-ip\cdot(x-y)}-e^{ip\cdot(x-y)}],\\ &&p_0=|\sqrt{(\overrightarrow{p}^2+M^2)}|.\nonumber
\end{eqnarray}
Hence,
\begin{equation}
[A^K_\mu(x)\, ,\, A^K_\nu(y)]=\frac{1}{\xi G M^2}\eta_{\mu\nu}D(x-y).
\end{equation}
This commutator is a multiple of the identity.

Now we have
%\begin{widetext}
\begin{eqnarray} \label{V}
V(J)=&\exp&\left\{-\int_{-\infty<t'\leq 0} d^4x'\int_{-\infty<t\leq 0} d^4x\ [A^K_\mu(x')J^\mu(x')\,,\,A^K_\nu(x)J^\nu(x)]\right\}\nonumber\\
&\times&\exp\left\{-i\int_{-\infty<t\leq 0} d^4x\ A^K_\mu(x)J^\mu(x)\right\}.
\end{eqnarray}
%\end{widetext}

The first factor does not contribute to (\ref{VAV}) as it is a multiple of identity where now $A_\mu=A^K_\mu$ so that
\begin{eqnarray}
&a')&V(J)^\dagger A_\mu^K(x) V(J)\nonumber\\
   &=&A^K_\mu(x)+\frac{i}{\xi G M^2}\int_{-\infty<t'<0}dx'\, J^\mu(x') D(x'-x).
\end{eqnarray}
We find, with $x_0=0$,
\begin{eqnarray}
&V&(J)^\dagger A^K_\mu(x) V(J) :=A_\mu^K+\Delta A_\mu^K\\
&=& A^K_\mu-\frac{P_\mu P_0/M}{\xi G M^2}\int \frac{d^3p}{2p_0(2\pi)^3}\frac{e^{-i\overrightarrow p\cdot\overrightarrow x}}{P\cdot p+i\epsilon}+c.c. \label{VAV2}
\end{eqnarray}

\subsection{ Calculation of (\ref{VeV})}
This term can be estimated from the commutator (\ref{Aecomm}). The first factor in (\ref{V}) does not contribute to (\ref{VeV}) while the second, in the Coulomb gauge, becomes our expression in \cite{Balachandran:2014hra,Balachandran:2015dva}, namely
%$$e^{-i\int d^3x A^K_i(x) \omega_i(x)}$$

$$e^{\int d^3 x ( A_i^- (x) \omega_i^+ (x) - A_i^+ (x) \omega_i^- (x))}$$
where $A_i^+$,$A_i^-$ are the positive and negative frequency parts of $A_i$ and 
\begin{equation}
\omega_i^{\pm} (x)=\mathcal O\left(\frac{1}{|\overrightarrow x|^2}\right)\ \textrm{as}\ |\overrightarrow x|\longrightarrow \infty.
\end{equation}
We do not need the detailed expression for $\omega_i^{\pm}$, which can be found in \cite{Balachandran:2015dva}.

Thus,  $e_i$ gets shifted by terms $\Delta e=\mathcal O\left(\frac{1}{r^2}\right)$. But $A$ as $r\longrightarrow\infty$ is $O\left(\frac{1}{r^2}\right)$. So $\Delta e\wedge A$ is $O\left(\frac{1}{r^4}\right)$ as $r\longrightarrow\infty$ and does not contribute to asymptotic $E_{ADM.}$

In contrast, $e$ is $\mathcal{O}(1)$ as $r\longrightarrow\infty$ while the shift $\Delta A^K$ of $A^K$ as $|\overrightarrow x|\longrightarrow\infty$ is $\mathcal{O}(1/r)$. This follows from (\ref{VAV2}) which gives $A^K(\lambda \overrightarrow x)=\frac{1}{\lambda }A^K(\overrightarrow x)$ (Change $p$ to $p'=\lambda p$ in $A^K(\lambda \overrightarrow x)$ to get this result.) Hence the term $e\wedge \Delta A^K$ term does contribute to the shift of energy by infrared terms. We estimate this shift next.

%In contrast, $e$ is $O(1)$ as $r \longrightarrow \infty$ while the shift $\Delta A^K$ of A^K as  $   |x| \longrightarrow \infty$ is $O(1/r)$. This follows from  (\ref{VAV2}) which gives $A^K (\Lambda x vectir) =1/\Lambda A^K ( xvector)$ ( Change $p$ to $p'= lambda p$ in $A^K(lambda x vector$ to get this result.) Hence  the term  $e \wedge \Delta A^K$ does contribute to the shift of energy by infrared gravitons. We estimate this shift next.

\section{ ADM Energy: The Discussion} We first calculate the shifts
\begin{eqnarray}
    \Delta E_W &=&  \lim_{r\to \infty} \int d\Omega \, \Delta \mathcal{E}_W \\
    \Delta \mathcal{E}_W &=& \frac{i}{2\pi G} \epsilon^{ijk} \hat{x}_i \bar{\psi} \gamma_5 \gamma_j \Delta A^K_k \psi
\end{eqnarray}
of the ADM energy $E_W$ and the energy density $\mathcal{E}_W$ caused by the infra-spin cloud.

\subsection{ Calculation of $\Delta \mathcal{E}_W$}

We have 
\begin{equation}
\Delta A^K_k = \Delta A_k \Sigma_{12}.
\end{equation}
With $\hat{x} = (0,0,1)$, that gives on simplification in the basis
\begin{eqnarray}
\gamma_\mu&=&\left( \begin{array}{cc}
\sigma_\mu & \phantom{m}\\
\phantom{m} & \tilde{\sigma_\mu}\end{array}\right),\\
\sigma_0&=&\mathbb{1},\ \sigma_i=-\tilde{\sigma_i}=\textrm{Pauli matrices},
\end{eqnarray}
\begin{eqnarray}
    \Delta \mathcal{E}_W &=& - \frac{M}{\xi (\sqrt{G}M)^4} r^2 \epsilon_{ijk} \chi_i P_j \hat{x}_k (I_++I_-=\bar I_+)\\
    \chi_i &=& \psi^{\dagger} \left( \begin{array}{cc}
\sigma_i \sigma_3 & 0  \\
0 & \sigma_i \sigma_3 \end{array} \right) \psi \\
    I_{+} &=& \int \frac{d^3 k}{(2\pi)^3 2k} e^{-i\vec{k}.\vec{x}} \frac{P_0}{P.k+i\epsilon}
\end{eqnarray}
We write 
\begin{equation}
\frac{P_0}{P.k+i\epsilon} = \frac{1}{k (1-\vec{v}.\hat{k})+i\epsilon}
\end{equation}
with $\vec{v} = \frac{P_i}{P_0}$ is the velocity of the particle. We integrate over $k$ to get
\begin{equation}
I_+ = \frac{-i}{r} \frac{1}{2(2\pi)^3} \int \frac{1}{\hat{k}.\hat{x} - i\epsilon} \frac{1}{1-\vec{v}.\hat{k}} d\Omega_k    
\end{equation}
where finally, $\epsilon \rightarrow 0+$. 

Setting $\hat{x} = (0,0,1)$ and using $\frac{1}{\hat{k}.\hat{x} - i\epsilon} = \frac{P}{\hat{k}.\hat{x}} + i\pi \delta (\hat{k}.\hat{x} = cos \theta_k)$, we integrate over $cos \theta_k$ to find 
\begin{equation}
I_+ + I_- = \frac{1}{2r (2\pi)^2} \int\limits_{0}^{2\pi} \frac{d\phi_k}{1-v_x cos\phi_k - v_y sin \phi_k}
\end{equation}
or with $z = e^{i\phi_k}$, $v_{\pm} = v_x \pm i v_y$,
\begin{eqnarray}
I_+ + I_- &=& \frac{1}{r (2\pi)^2 } \frac{i}{v_-} \oint\limits_{|z|=1} \frac{dz}{(z-z_+)(z-z_{-})},\label{integral}\\
z_{\pm} &=& \frac{1}{v_-} \pm \frac{1}{v_-} \sqrt{1-v_+ v_-}.
\end{eqnarray}
We note that 
\begin{eqnarray}
|z_+ z_-| &=& |\frac{v_+}{v_-}| = 1, \\ 
z_+ \bar{z}_+ &=& \frac{1}{|\vec{v}|^2} + \left( 1- \frac{1}{|v|^2} \right)^2 > 1 \, \text{as} \, |\vec{v}|^2 < 1 
\end{eqnarray}
so that the integral (\ref{integral}) is well-defined and encloses only the pole at $z_-$. Hence 
\begin{eqnarray}
I_+ + I_- &=& \frac{1}{r (4\pi)} \frac{1}{\sqrt{1- v_+ v_- }} \\
&=& \frac{1}{r (4\pi)} \frac{1}{\sqrt{1-|\vec{v}|^2 + (\vec{v}.\hat{x})^2}}
\end{eqnarray}
and finally
\begin{equation}
\Delta \mathcal{E}_W = - \frac{1}{4 \pi \xi} \frac{Mr}{(\sqrt{G}M)^4} \frac{1}{\sqrt{1-|\vec{v}|^2 + (\vec{v}.\hat{x})^2}} \epsilon_{ijk} \chi_i P_j \hat{x}_k
\end{equation}

\subsubsection{ $\Delta E_W = 0$}
This result follows from rotational invariance. Thus
\begin{equation}
\int d\Omega _{\hat{x}_k} \frac{\hat{x}_k}{\sqrt{1-|\vec{v}|^2 + (\vec{v}.\hat{x})^2 }}
\end{equation}
is proportional to $\frac{P_k}{P_0}$ and hence $\Delta E_W= 0$. 

\subsubsection{ $\Delta \mathcal{E}_W$}
From $\Delta E_W = 0$, we see that $\Delta \mathcal{E}_W$ is not uniformly distributed over angles. Let us look for the maxima of $|\Delta \mathcal{E}_W|$. 

The expression $\chi_i$ refers to the direction of momentum of gravity without source ( eq. (56) in \cite{Witten:1981mf}). A simple choice for $\psi$ which gives the maximum of $|\chi_3|$ is 
\begin{equation}
\psi = \frac{1}{\sqrt{2}}\begin{pmatrix}1\\0\\1\\0\end{pmatrix}
\end{equation}
where also $\chi_{1,2}$ are zero. In that case if $\vec{P} = P(1,0,0)$, $\hat{x} = (0,1,0)$, we have $\epsilon_{ijk} \chi_i P_j \hat{x}_k = P$.

There is also the cut-off $r$. If we take the limit $r\rightarrow \infty$, $\Delta \mathcal{E}_W \rightarrow \infty$. If at the same time, $\Delta E_W \rightarrow \infty$, we can say that the generator of time translation, i.e. the ADM Hamiltonian, is spontaneously broken. But in fact, $\Delta E_W = 0$. So the divergence of $\Delta \mathcal{E}_W$ needs clarification. Tentatively, we take $r$ to be Compton wavelength $1/M$ of the particle and set $Mr = 1$. 

We now eliminate the unknown $\xi$ by taking the ratio of energy density shifts $\Delta \mathcal{E}_W^m$ and $\Delta \mathcal{E}_W^M$ for velocities $\vec{v}_m$ and $\vec{v}_M$ to get 
\begin{equation}
    \frac{\Delta \mathcal{E}^m_W}{\Delta \mathcal{E}^M_W} \approx \frac{M^4 \sqrt{(1-|\vec{v_M}|^2)}}{m^4 \sqrt{(1-|\vec{v_m}|^2)}}\label{ratio}
\end{equation}
This shows that faster particles of smaller mass create greater shift of energy density. 

Note that in the ratio (\ref{ratio}), we get the ratio of the  masses if the cut-offs are same or proportional. That is finite even if the cut-ff $r$ is allowed to approach infinity. 

We do not have a good way to evaluate $\xi$ and hence cannot report and estimate of $\Delta \mathcal{E}_W$. But it seems probable that the spin cloud will contribute to $\Delta \mathcal{E}_W$ in any model and that it will diverge like $r$ as $r \rightarrow \infty$.

Finally we remark that in QED and QCD, the infra-red cloud of photons or gluons causes spontaneous breakdown of Lorentz symmetry. This happens because the boast operators can not be unitarily implemented on the in-states which are dressed as in (\ref{instate}). We expect similar results to hold in case of gravity. However, since the asymptotic Lorentz group does not act on general spacetimes globally, it is not clear how to write the boast generators prior to infrared effects in general.

\section{Acknowledgements}

Rakesh Tiberwala derived eq. (\ref{Aecomm}) for A.P.B. Alok Laddha and T.R.Govindarajan kindly arranged his visit to Chennai Mathematical Institute. He thanks them for intellectual and material support.

\end{document}